\documentclass[
  twocolumn,
  prb,
  showpacs,
  amsmath,
  amssymb,
  superscriptaddress
]{revtex4}

\usepackage{bm}
\usepackage{bbm}
\usepackage{graphicx}

\newcommand{\sh}{\mathop{\mathrm{sh}}}
\newcommand{\ch}{\mathop{\mathrm{ch}}}
\begin{document}

\title{
Dissipationless kinetics of one dimensional interacting fermions}

\author{I.\ V.\ Protopopov}
\affiliation{
 Institut f\"ur Theorie der Kondensierten Materie and DFG Center for Functional
Nanostructures, Karlsruhe Institute of Technology, 76128 Karlsruhe, Germany
}
\affiliation{
 L.\ D.\ Landau Institute for Theoretical Physics RAS,
 119334 Moscow, Russia
}

\author{D.\ B.\ Gutman}
\affiliation{Department of Physics, Bar Ilan University, Ramat Gan 52900,
Israel }
\affiliation{
 Institut f\"ur Nanotechnologie, Karlsruhe Institute of Technology,
 76021 Karlsruhe, Germany
}

\author{M.\ Oldenburg}
\affiliation{
 Institut f\"ur Theorie der Kondensierten Materie and DFG Center for Functional
Nanostructures, Karlsruhe Institute of Technology, 76128 Karlsruhe, Germany
}

\author{A.\ D.\ Mirlin}

\affiliation{
 Institut f\"ur Nanotechnologie, Karlsruhe Institute of Technology,
 76021 Karlsruhe, Germany
}
\affiliation{
 Institut f\"ur Theorie der Kondensierten Materie and DFG Center for Functional
Nanostructures,
 Karlsruhe Institute of Technology, 76128 Karlsruhe, Germany
}


\affiliation{
 Petersburg Nuclear Physics Institute,  188300 St.~Petersburg, Russia.
}

\begin{abstract}
We study the problem of evolution of a density pulse of one-dimensional 
interacting fermions with a non-linear single-particle spectrum.
 We show that, despite non-Fermi-liquid nature of the problem, 
non-equilibrium phenomena can be described in terms of a kinetic equation
for certain quasiparticles related to the original fermions by a non-linear
transformation which decouples the left- and right-moving excitations. 
Employing this approach, we investigate the kinetics of the phase space
distribution of the quasiparticles and thus determine the time evolution of the
density pulse. This allows us to explore a crossover from the essentially
free-fermion evolution for weak or short-range interaction to hydrodynamics
emerging in the case of sufficiently strong, long-range interaction. 
\end{abstract}


\maketitle
Understanding non-equilibrium phenomena  is one of central
themes in condensed matter physics. For Fermi-liquid  systems 
(e.g. electrons in metals) such phenomena are
conventionally described in the framework of a quantum kinetic equation for
quasiparticle excitations. According to Landau Fermi-liquid theory, it has the same form as for weakly interacting particles up to  a
renormalization of parameters (effective mass, interaction constants, and
scattering integral). This equation governs the  evolution of a
single-particle density matrix (characterizing the quasiparticle phase
space distribution) and readily yields various physical observables \cite{Landau1,Landau2,Kadanoff}.

For a variety of strongly interacting fermionic systems, the Fermi
liquid theory (at least, in its standard form) is not applicable: interaction
destroys the quasiparticle pole. In these cases on has to find an alternative
way to describe transport and non-equilibrium phenomena. This is usually done by
formulating effective theories in terms of some collective degrees of freedom. 
A famous realization of a non-Fermi-liquid
state is provided by one-dimensional (1D) interacting fermions. 
This system is characterized by a strongly correlated ground state---Luttinger
liquid (LL)\cite{Stone_book,Delft,Gogolin,Maslov,Giamarchi}---which exhibits  
an infrared divergence of an electronic self-energy, eliminating the
quasiparticle pole from the spectral function. This manifests itself in a
power-law suppression of the tunneling (zero-bias anomaly) and indicates
that quasiparticle excitations are ill-defined. A well-known tool for dealing
with such correlated 1D systems is bosonization
\cite{Stone_book,Delft,Gogolin,Maslov,Giamarchi}. After linearization of the
fermionic spectrum, it allows one to map the problem onto one of
non-interacting bosons. For arbitrary distribution functions, the
non-equilibrium bosonization yields results for  LL correlation functions in
terms of singular Fredholm determinants \cite{GGM,PGM}.   

In this work we explore kinetics of interacting 1D fermions,
having in mind the following model setup. Initially, a hump (or a dip) in a
fermionic density is created by an external potential. At time $t=0$ the
potential is switched off, and electronic pulses start to propagate to the right and to the
left. The evolution of the electronic density as a function of time is
measured. While experiments of this type are particularly natural in the context
of cold atomic gases \cite{Bloch,Lahaye}, we expect them to be feasible also for
electronic systems. Since for a linearized spectrum the pulse moves without
changing its form, a curvature of the single-particle spectrum  is absolutely
essential  for the problem under consideration. Specifically, the curvature
induces a tendency to an ``overturn'' of the pulse at a certain time $t_c$, thus
making the pulse evolution for times $t>t_c$ a challenging problem
\cite{bettelheim06}.

The non-linearity of a fermionic spectrum induces an interaction between 
bosonic collective modes
\cite{Haldane81,schick68,Jevicki_Sakita,Sakita,Polychronakos,abanov05,Stone}, 
giving rise to a quantum hydrodynamic theory. Such ``non-linear Luttinger
liquids'' arise in a variety of fermionic, bosonic, and spin
system and have recently attracted a considerable attention 
\cite{imambekov09,imambekov11}. 

A natural idea is to try to tackle the interaction between the bosonic modes
perturbatively\cite{Aristov}. As it turns out,  the 1D character of the problem
induces infrared singularities invalidating the naive perturbative expansion. 
The bosonized theory is treatable only in the limit of
strong and long-ranged interaction, which justifies the  saddle-point
approximation, as was done in
Ref.~\cite{bettelheim06} for Calogero model and in Ref.~\cite{PGM2012} for a
generic interaction. Equations of motion obtained in this way  can be
viewed as  Euler and continuity  equation for  an ideal fluid,  and therefore the system
is described by a non-dissipative classical hydrodynamics. Depending on the sign
of the initial pulse, an interplay between non-linearity and dispersion  leads to  emergence of strong density oscillations or of  solitons
after the shock \cite{PGM2012}.

The problem has been also studied  in the opposite limit of free fermions
\cite{PGM2012,Glazman_Bettlheim}, where the evolution of 
Wigner function is described by a simple kinetic
equation. For sufficiently long times, $t>t_c$, a population inversion occurs,
leading to density oscillations that can be viewed as Friedel-type oscillations
between different Fermi edges. 

Thus, the pulse evolution was analyzed in two opposite limits 
(no vs.   strong long-range interaction) by  different means (fermionic
vs. bosonic),  and within  different physical pictures (inverted
population vs. hydrodynamic waves). We now address this
problem for an arbitrary  interaction. 
By bosonizing the system, performing a certain unitary
transformation and refermionizing it,
we explicitly build corresponding quasiparticle operators  and
formulate a kinetic description in their terms. The latter
describes, in particular, the sought density evolution.

The problem is characterized by a Hamiltonian $H=H_0+H_{\rm int}$, where
the kinetic part $H_0$ describes two spinless chiral modes
(labeled by subscript $\eta=R, L$ or, occasionally, $\eta=\pm 1$) with a
non-linear spectrum 
\begin{equation}
H_0=\sum_{\eta, k}\eta k v_F :a_{\eta k}^+a_{\eta k}:+ (1/2m)\sum_{\eta,
k}k^2 :a^+_{\eta k}a_{\eta k}:\,. 
\end{equation}
The interaction part reads
\begin{equation}
H_{\rm int}=(1/2)\int dx_1 dx_2 g(x_1-x_2)\rho(x_1)\rho(x_2)\,,
\end{equation}
where $\rho = \rho_L + \rho_R$ is the density.
The kinetic term can be bosonized 
as follows \cite{schick68}
\begin{equation}
H_0=\pi v_F \int dx \left( \rho_R^2+\rho_L^2\right)+(4\pi^2/6m) \int\left(\rho_R^3+\rho_L^3\right)
\end{equation}
with Fourier components of the densities satisfying the standard
commutation relations ($L$ is the system length)
$[\rho_{\eta,
q},\rho_{\eta',-q'}]=\eta\delta_{\eta,\eta'}\delta_{q,q'}Lq/2\pi\,$.
The interaction mixes the chiral sectors. 
On the quadratic level, this coupling can be eliminated by a
canonical transformation $R_q=U_2\rho_{R,q}U_2^\dagger$,
$L_q=U_2\rho_{L,q}U_2^\dagger$ of the standard Bogoliubov form  
\begin{eqnarray}&&
\rho_{R, q}=\cosh \kappa_q R_q-\sinh\kappa_q L_q\,, \\&& 
\rho_{L, q}=-\sinh \kappa_q R_q +\cosh\kappa_q L_q, 
\end{eqnarray}
where
$\tanh 2\kappa_q= g_q/(2\pi v_F+g_q)$.
In terms of new fields, the quadratic part is
\begin{equation}
H^{(2)}=(\pi/L)\sum_{q}u_q\left(R_qR_{-q}+L_qL_{-q}\right)\,,
\end{equation}
with a sound velocity $u_q=v_F (1+ g_q/\pi v_F)^{1/2} = v_F/K_q$.
\label{sound_velocity}

As a side effect of Bogoliubov transformation, the cubic part of the
Hamiltonian 
acquires a form that mixes the right and left movers:
\begin{eqnarray}
H^{(3)}&=& (2\pi^2/3m L^2) \sum_{{\bf q}}
\Gamma_{\bf q} \left[ (R_1R_2R_3+L_1L_2L_3 ) \right. \nonumber \\&
+ & \left. 3\Gamma'_{\bf q} (R_1R_2L_3+L_1L_2R_3)\right].
\end{eqnarray}
Here we have introduced notations ${\bf q}\equiv \{q_1,q_2,q_3\}$,
$R_i=R_{q_i}$,  $L_i=L_{q_i}$; the  summation over ${\bf q}$ is restricted to
$q_1+q_2+q_3=0$ and we have  defined vertices ($\kappa_i \equiv
\kappa_{q_i}$) 
\begin{eqnarray}&&
\Gamma_{\bf q}=\ch \kappa_1\ch \kappa_2 \ch \kappa_3-\sh \kappa_1\sh \kappa_2\sh \kappa_3, \nonumber  \\&&
\Gamma'_{\bf q}=\sh \kappa_1\sh \kappa_2\ch \kappa_3 -\ch \kappa_1\ch
\kappa_2\sh \kappa_3\,.
\end{eqnarray}
The decoupling of the right and left  sectors of the theory can be extended to
the cubic level. To this end, we  perform an  additional  unitary transformation
$\tilde{\rho}_R= U_3 R U_3^\dagger$ and 
$\tilde{\rho}_L= U_3 L U_3^\dagger$,
determined by the operator 
\begin{eqnarray}
U_3=\exp \sum_{{\bf q}}[f_{\bf q} R_1 R_2 L_3 - (L\leftrightarrow R)],
\end{eqnarray}
where 
\begin{equation}
f_{\bf q}=\frac{2\pi^2}{mL^2}
\frac{\Gamma^{'}_{\bf q}}{u_{q_1} q_1+u_{q_2} q_2-u_{q_3} q_3}\,.
\end{equation}
After this transformation, the Hamiltonian $H$ mixes the left and right modes
only due to the terms quartic in  the density
\begin{eqnarray}
\label{Hamiltonian_bos}
H &=& (\pi/L) \sum_{\eta,
q}u_q\tilde{\rho}_{\eta,q}\tilde{\rho}_{\eta,-q}\nonumber \\&
+ & (2\pi^2/ 3m L^2)\sum_{\eta,{\bf q}}
\Gamma_{\bf q} \tilde{\rho}_{\eta,1}\tilde{\rho}_{\eta,2}\tilde{\rho}_{\eta,3}
+O(\tilde{\rho}^4)\,.
\end{eqnarray}
One can continue the procedure described above to disentangle the left and right
movers order by order in perturbation theory in $\rho/mv_F$.
This allows us to decouple the Hamiltonian into chiral sectors with an
arbitrary accuracy. For our purposes,  the transformations $U_2$ and $U_3$ are  
sufficient, and terms containing four and more density operators  will 
be neglected.

We have thus obtained a chiral bosonic theory (\ref{Hamiltonian_bos}), with
interaction originating from the non-linearity of the fermionic spectrum
and a $q$-dependent sound velocity originating from the electron-electron 
interaction. We now proceed  by refermionizing this theory,
following the idea put forward in Ref.~\cite{Rozhkov} (see also \cite{imambekov09}), 
where such a mapping was performed after the conventional Bogoliubov transformation $U_2$. It is crucial
for our problem that we also  carry out the transformation $U_3$, decoupling the
chiral  sectors, and only then refermionize. 
More specifically, we define ``composite fermion'' operators
that are built from the original ones by  consecutive rotations  
\begin{equation}
\tilde{\Psi}_\eta=U_3U_2 \Psi_\eta U^\dagger_2U^\dagger_3\,.
\end{equation}
Since the rotation is exponential in the density fields, this  somewhat
resembles the composite-fermion transformation in the fractional quantum Hall
regime.
In terms of the new operators, the  Hamiltonian is given by
\begin{eqnarray}
\label{Hamiltonian_ref}
H &=& \sum_{\eta,k}\tilde{\Psi}^\dagger_{\eta,k}\left(\eta u_0k
-\frac{k^2}{2m^*} \right) \tilde{\Psi}_{\eta,k} 
+\frac{1}{2L} \sum_{\eta,q}  V_q \tilde{\rho}_{\eta,q}\tilde{\rho}_{\eta,-q}
\nonumber \\&
+&\frac{2\pi^2}{3m L^2}\sum_{\eta,{\bf q}} \gamma_{\bf
q}\tilde{\rho}_{\eta,q_1}\tilde{\rho}_{\eta,q_2}\tilde{\rho}_{\eta,q_3}\,.
\end{eqnarray}
The quadratic part of the Hamiltonian
({\ref{Hamiltonian_ref}) is parametrized by the renormalized  Fermi velocity
$u_0\equiv u_{q=0}$ and  the spectral curvature
\begin{equation}
1/m^* \simeq \Gamma_{{\bf q}=0} / m.  
\end{equation}
There is also a residual interaction between particles represented by
two-particle and three-particle vertices
\begin{equation}
V_q=2\pi(u_q-u_0)\,, \qquad \gamma_{\bf q}=\Gamma_{\bf q}-\Gamma_{{\bf q}=0}.
\end{equation}
The residual interaction $V_q$  vanishes  at low  momenta, 
($V_q \propto  q^2$   for a generic finite-range interaction) and is irrelevant
in the renormalization-group sense. 
The three body interaction is still weaker  ($\gamma_{\bf q} \propto   {\bf
q}^2$ and, in addition, contains the  factor $\rho/m^*u \ll 1$) and  we  neglect it from now
on \cite{3-body-interaction}.
The disappearance of the interaction at small momenta, 
makes perturbation theory for the  composite fermions regular
in the infrared limit, and  the system behaves as a weakly
interacting Fermi gas.

We define the quasiparticle density matrix 
\begin{eqnarray}
\tilde{f}_\eta(x,y,t) &=& \langle
\tilde{\Psi}_\eta^\dagger(x-y/2,t)\tilde{\Psi}_\eta(x+y/2,
t)\rangle \nonumber \\
&=& \int (dp/2\pi)e^{ipy}f_\eta(x,p,t) 
\end{eqnarray}
 that  within the Hartree approximation satisfies the collisionless quantum
kinetic equation 
\begin{eqnarray}
\label{Vlasov}
&& \partial_t \tilde{f}_\eta(p,x,t) + (p/m^*)\partial _x \tilde{f}_\eta(p,x,t)
+\int (dp/2\pi) e^{-ipy} \nonumber \\
&&\times \tilde{f}_\eta(x,y,t)
[\tilde{\phi}_\eta(x+y/2)-\tilde{\phi}_\eta(x-y/2)]=0\,,
\end{eqnarray}
with the self-consistent electric field
\begin{equation}
\tilde{\phi}_\eta(x,t)=\int dx'  V(x-x')\tilde{\rho}_\eta(x',t),\,.
\end{equation}
To obtain the physical density $\rho$ out of the solution $\tilde{\rho}$ one needs to  use  the 
relation between the densities; in the leading order 
$\rho\simeq \sqrt{K} \tilde{\rho}$, Appendix  \ref{Supplementary1}.
Note that Eq. (\ref{Vlasov}) is exact in the limits of non-interacting
electrons and of a harmonic LL ($m\rightarrow
\infty$, arbitrary electron interaction), see Appendix \ref{Supplementary2}.

In order to analyze the pulse dynamics, we solve Eq.(\ref{Vlasov}) numerically (see Appendix \ref{Supplementary3}), focusing
on times exceeding the "shock formation time" $t_c$ when the phase space distribution of
non-interacting fermions develops an inverse population. For initial density
perturbation of the amplitude $\Delta\rho$ and spacial extent $\Delta x$ one finds
$t_c\sim m\Delta x/\Delta\rho$.  
The Wigner function in the initial state  was discussed in
Refs.~\cite{Bettelheim2011, PGM2012}, see also Appendix \ref{Supplementary2}. We plot it
in Fig.\ref{fig1} for  a gaussian density hump 
($\tilde{\rho_0}(x)=\Delta\rho\exp(-x^2/2\sigma^2)$ with  $\sigma=200/mv_F$ and $\Delta\rho =0.01 mv_F$)  in the  initial state.
Besides changing from $0$ to $1$ at classical Fermi surface
$p_F(x)=2\pi\widetilde{\rho}_0(x)$, the Wigner function  
exhibits phase-space oscillations (that do not manifest
themselves in the total density for a spatially smooth hump).

\begin{figure}
\includegraphics[width=8.6cm]{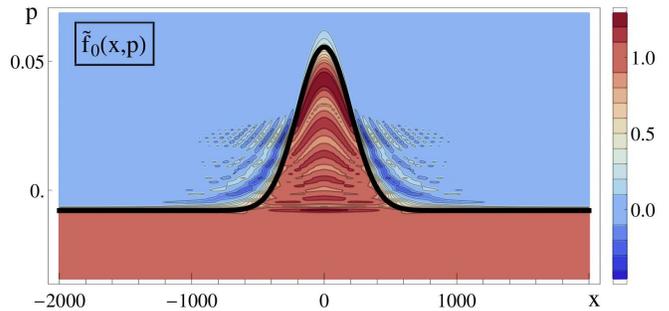}
\caption{\small The initial quasiparticle Wigner function $\widetilde{f}_0(x,
p)$. Thick black line shows the classical Fermi surface $p_F(x)=2\pi\rho_0(x)$. 
}
\label{fig1}
\end{figure}


While our approach is very general, we now focus on a model of
finite range interaction  
\begin{equation}
g(q)= (1/l_0m) \exp (-q^2l_{\rm int}^2)\,,
\end{equation}
with two lengths $l_0$ and $l_{\rm int}$ parameterizing its strength and range.
The classical hydrodynamics emerges if two conditions are fulfilled
\begin{equation}
 \label{hydrodyn-conditions}
 l_0\Delta\rho \ll 1 \,, \qquad l_{\rm int}^2\Delta\rho / l_0 \gg 1 \,.
\end{equation}
In the opposite limit (if at least one of the inequalities $l_0\Delta\rho \gg 1$
and $l_{\rm int}^2\Delta\rho / l_0 \ll 1$ is fulfilled)  the solution remains
close to that for free fermions. To illustrate the behavior of the solution of
the kinetic equation (\ref{Vlasov}) in both regimes and in a crossover between
them, we fix $l_0 = 1/mv_F$ and $\Delta\rho=0.01mv_F$  such that the first of
the conditions (\ref{hydrodyn-conditions}) is well fulfilled and vary $l_{\rm
int}$.

\begin{figure}
\includegraphics[width=8.6cm]{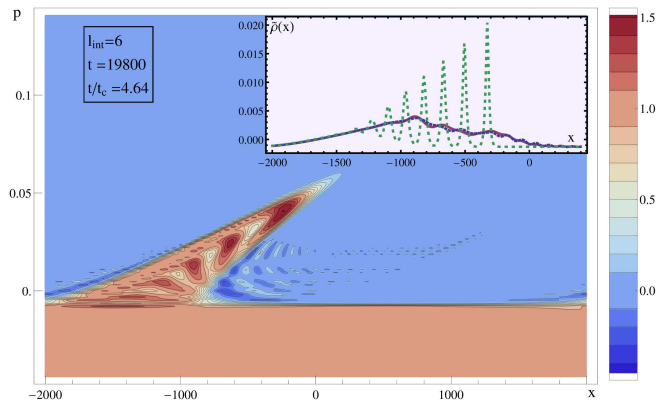}
\caption{\small Quasiparticle phase-space distribution (Wigner function) for a
short interaction range, $l_{\rm int}=6/mv_F$, at $t=4.6t_c$.
Inset: corresponding density (solid red line) in comparison to the
density of non-interacting fermions (green dotted) and the
predictions of classical hydrodynamic theory (dashed blue).}
\label{fig2}
\end{figure}

For a sufficiently short-range  interaction, an inverse population develops
for $t> t_c$. This is demonstrated in Fig. \ref{fig2}, where a  snapshot of the
phase space at time $t=4.64 t_c$ is shown  for interaction range
$l_{\rm int}=6/mv_F$. In this case the second parameter 
of Eq.~(\ref{hydrodyn-conditions}) is relatively small, i.e. $l_{\rm
int}^2\Delta\rho/l_0 = 0.36$. 
The inset shows the corresponding density in comparison
to that of non-interacting fermions and the
predictions of hydrodynamic theory. As one sees, the interacting density is close
to that of free fermions, meaning that  the ``composite fermion'' interaction
effects are weak, as expected. It should be emphasized that the original
electron interaction may well be strong in this regime; i.e. the parameter
$1/l_0 mv_F$ does not need to be small. (In our modeling it is equal to unity
and can also be larger.)
As for free fermions, one observes oscillations
of the total density that originate from the phase-space oscillations in the
initial state and develop  in the region where  the inverse population is
formed \cite{PGM2012,Glazman_Bettlheim}.
We also provide a comparison with the density calculated
by using a classical hydrodynamic equation (obtained as a saddle-point of the
bosonic theory). Clearly, the classical hydrodynamics, which
yields much stronger oscillations, is not a proper way to describe
the system in this regime of weakly-interacting quasiparticles. 

As the quasiparticle interaction becomes stronger ($l_{\rm int}=20/mv_F$), the
density significantly deviates  from the free fermion limit and
the agreement with the hydrodynamics improves, see Fig.\ref{fig3}.
However, the system still shows clear traces of the population
inversion leading to deviations from the hydrodynamic solution that
proliferate with time and become quite substantial at $t=4.6t_c$.
In this intermediate regime neither free-fermion model nor
hydrodynamic approximations are valid, and  the kinetic approach  is the only
adequate tool to controllably address the problem. 

\begin{figure}
\includegraphics[width=8.6cm]{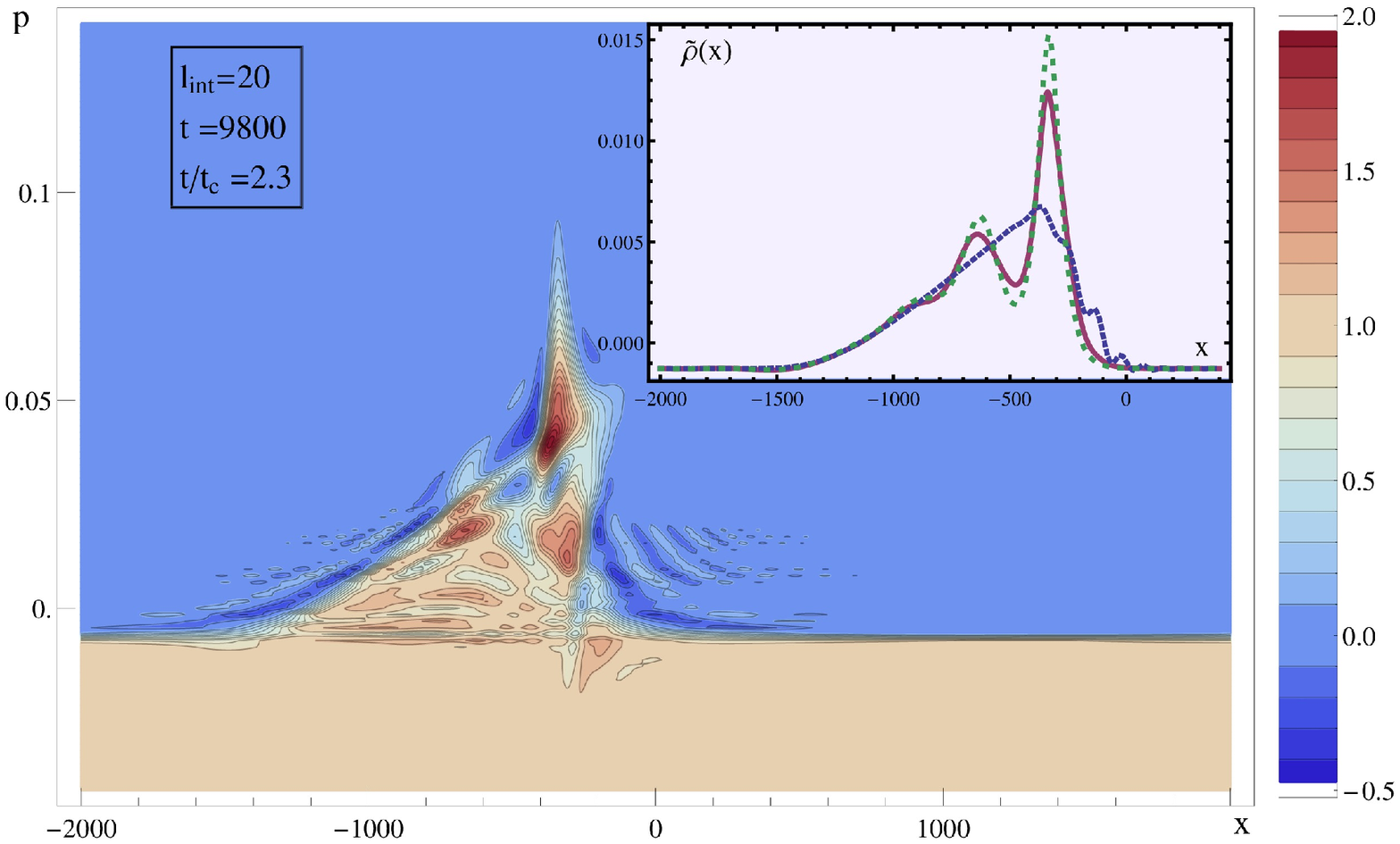} \\[0.2cm]
\includegraphics[width=8.6cm]{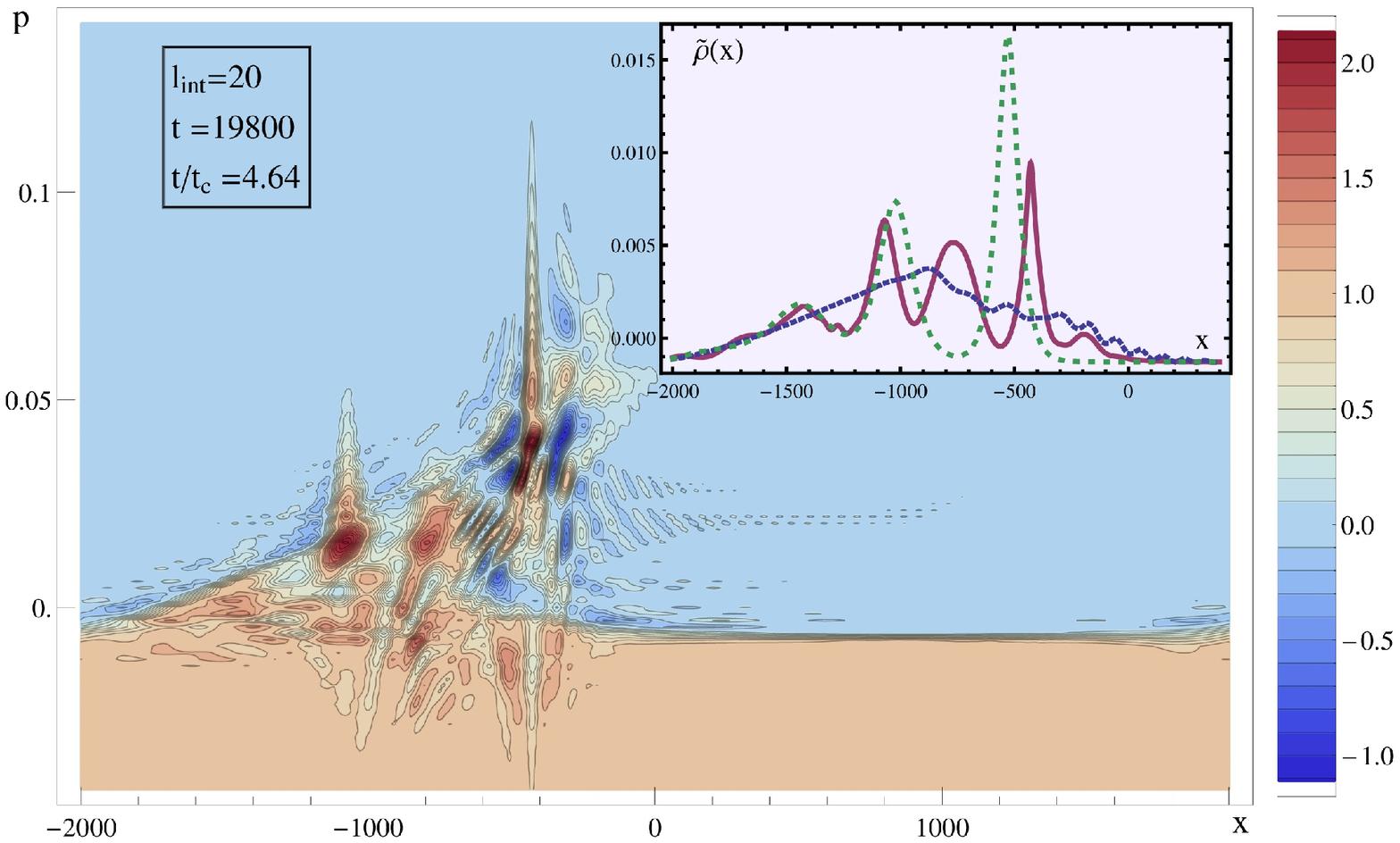}
\caption{\small Same as in Fig.~\ref{fig2} but for a medium-range
interaction, $l_{\rm int}=20/mv_F$, at $t= 2.3t_c$ (top) and $t=4.64t_c$
(bottom).}
\label{fig3}
\end{figure}


With a further increase of the interaction strength ($l_{\rm int}=20/mv_F$)
the agreement between hydrodynamic and kinetic approaches is reached, see
Fig.~\ref{fig5}. In this regime the phase space distribution is approximately
given by a Fermi function with a position-dependent Fermi momentum
$p_F(x)$,  determined by the classical hydrodynamic
equation. On top of sharp Fermi surface, we observe an additional ``fine structure'' in the
phase-space distribution,  shown in Fig.~\ref{fig5}. It remains to be seen
whether these details of the quantum state,  which are beyond the hydrodynamic
picture,  lead to strong deviations from the hydrodynamic solution at  the longer  times.

\begin{figure}
\includegraphics[width=8.6cm]{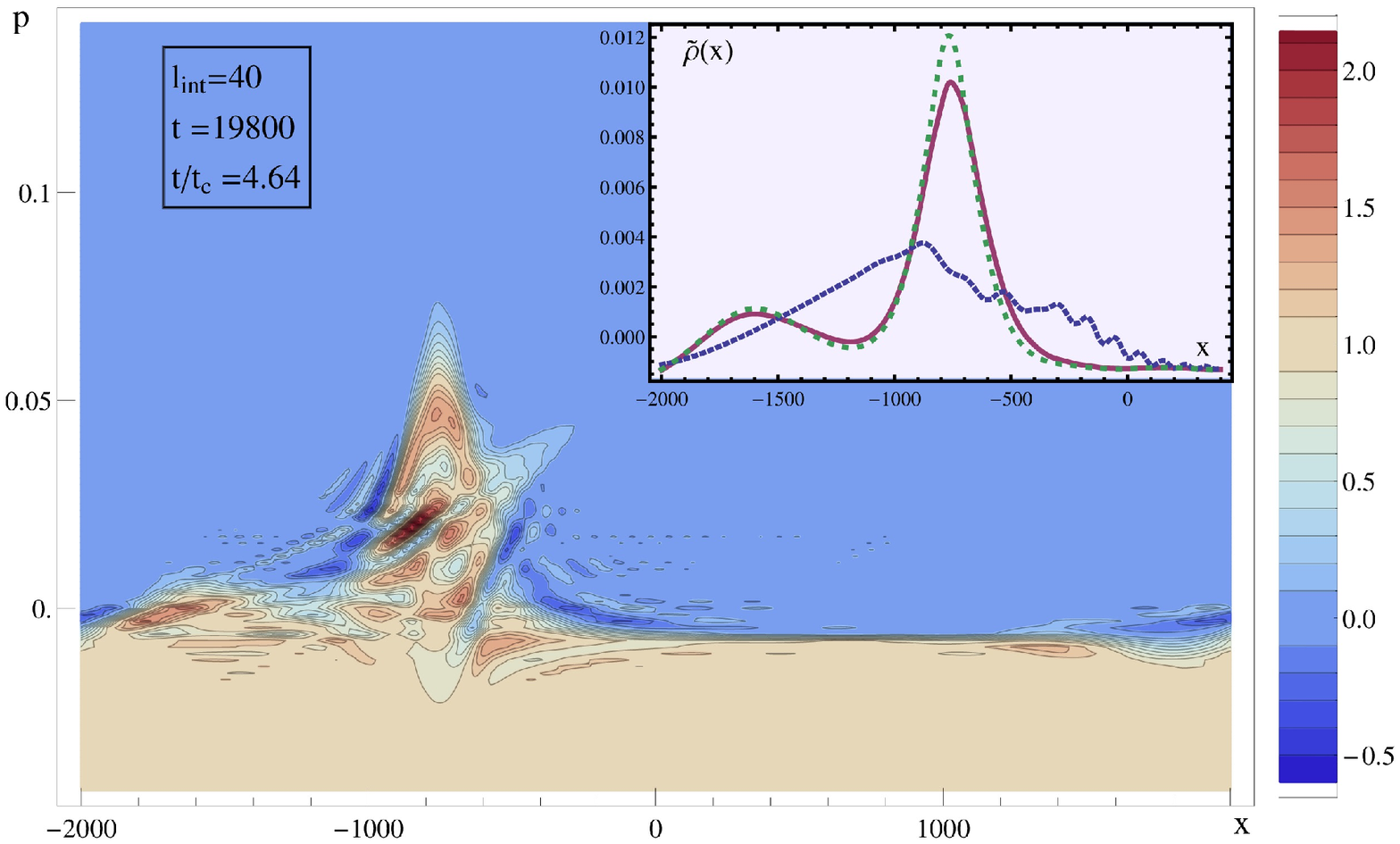}
\caption{\small Same as in Fig.~\ref{fig2} but for a long-range
interaction, $l_{\rm int}=40/mv_F$, at time $t=4.6t_c$.}
\label{fig5}
\end{figure}

In addition to  the selfconsistent electric field,  the
quasiparticle interaction in Eq.~(\ref{Hamiltonian_ref}) causes inelastic
quasiparticle scattering. When taken into account, these processes generate
a collision integral in the kinetic equation (\ref{Vlasov}). Dominant
contributions originate from triple collisions
\cite{khodas2007,Lunde2007,Karzig2010,Matveev,Levchenko,Dmitriev} and from the $\tilde{\rho}^3$ term in
Eq.~(\ref{Hamiltonian_ref}).
A quick estimate shows that the rate $1/\tau_{\rm in}$ of such
processes is proportional to a high-power of a small parameter
$\Delta\rho/mv_F$ (or of $T/mv_F$ at finite temperature $T$) and is thus very
small. Therefore there is  a parametrically  broad range of times, $t < \tau_{\rm
in}$,  for which the collisionless kinetic equation studied in this work is
applicable. A detailed analysis of the inelastic relaxation leading to
a viscous hydrodynamics at $t > \tau_{\rm in}$ will be presented elsewhere. 

To summarize, we have studied  evolution of a density pulse of 1D 
interacting fermions with a non-linear single-particle spectrum.
We identified excitations that  play a role of weakly interacting 
quasiparticles for non-equilibrium phenomena inside the wire and 
described their  dynamics by a quantum kinetic equation.
The evolution of the corresponding phase space distribution is determined by 
two competing effects: the dispersion that  tends to overturn
Fermi surface, and the quasiparticle interaction that tends to stabilize it. 
Solving numerically the kinetic equation, we have demonstrated a crossover from 
the free-fermion-like evolution for weak or short-range interaction to
hydrodynamics emerging in the case of sufficiently strong, long-range
interaction.

Our work shows that while 1D interacting systems are not Fermi liquids in the
conventional sense, kinetic phenomena in such systems can be cast into 
Landau paradigm of weakly interacting fermionic quasiparticles. We foresee
numerous extensions and applications of our formalism, including
other types of interaction, relaxation
phenomena (also in presence of disorder), and edge states of integer and
fractional quantum Hall systems and topological insulators.

We acknowledge discussions with I.V.Gornyi and support by Alexander
von Humboldt Foundation, ISF, and GIF. 

\appendix
\widetext

\section{Physical density vs. density of composite fermions}
\label{Supplementary1}
In this section we write down  explicit expressions for the physical densities $\rho_{\eta}$ in terms of the densities of composite fermions.
The decoupling of the left and right-movers in the quadratic Hamiltonian $H^{(2)}$ (see Eq.(6) of the main text) is achieved via the Bogolubov transformation
\begin{eqnarray}&&
\label{Supp:rho_R}
\rho_{R, q}=U_2^+R_qU_2=\cosh \kappa_q R_q-\sinh\kappa_q L_q\,, \\&& 
\rho_{L, q}=U_2^+L_qU_2=-\sinh \kappa_q R_q +\cosh\kappa_q L_q\,.
\end{eqnarray}
To perform decoupling of the cubic terms one needs to perform the non-linear rotation
\begin{equation}
 R=U^+_3\tilde{\rho}_RU_3\\, \qquad L=U^+_3\tilde{\rho}_LU_3\,,
\end{equation}
with
\begin{eqnarray}
U_3=\exp\left(\sum_{{\bf q}}f_{\bf q} R_1 R_2 L_3 - (L\leftrightarrow R)\right)\,,
\\ f_{\bf q}=\frac{2\pi^2}{mL^2}
\frac{\Gamma^{'}_{\bf q}}{u_1 q_1+u_2 q_2-u_3 q_3}\,.
\end{eqnarray}
To third order in densities we obtain
\begin{eqnarray}
\label{Supp:R_rho_tilde}
 R_q=\tilde{\rho}_{Rq}+\frac{q L}{\pi}\sum_{2+3-q=0}f_{(-q, 2, 3)}\tilde{\rho}_{R2}\tilde{\rho}_{L3}-
\frac{q L}{2\pi}\sum_{1+2-q=0}f_{(1, 2, -q)}\tilde{\rho}_{L1}\tilde{\rho}_{L2}\,,\\
\label{Supp:L_rho_tilde}
L_q=\tilde{\rho}_{Lq}+\frac{q L}{\pi}\sum_{2+3-q=0}f_{(-q, 2, 3)}\tilde{\rho}_{L2}\tilde{\rho}_{R3}-
\frac{q L}{2\pi}\sum_{1+2-q=0}f_{(1, 2, -q)}\tilde{\rho}_{R1}\tilde{\rho}_{R2}\,.
\end{eqnarray}
The connection of $\rho_\eta$ and $\tilde{\rho}_\eta$ can be now read off from (\ref{Supp:rho_R}), (\ref{Supp:R_rho_tilde}) and (\ref{Supp:L_rho_tilde}).

The consideration above simplifies considerably when the relevant spacial scale of the density variation is small compared to the interaction radius. 
In this case the transformations $U_2$ and $U_3$ act locally in space leading to
\begin{eqnarray}
\rho_R(x)=\frac{\sqrt{K_0}}{2}(R(x)+L(x))+\frac{1}{2\sqrt{K_0}}(R(x)-L(x))\,,\\
\rho_L(x)=\frac{\sqrt{K_0}}{2}(R(x)+L(x))-\frac{1}{2\sqrt{K_0}}(R(x)-L(x))\,,
 \end{eqnarray}
and
\begin{eqnarray}
R(x)=
 \tilde{\rho}_R(x)+\frac{\pi}{m}\frac{1-K_0^2}{8u_0\sqrt{K_0}}
\left[-\frac{1}{\pi}\partial_x(\tilde{\rho}_R(x)\tilde{\varphi}_L(x))+\tilde{\rho}_L^2(x)\right]\,,\\
L(x)=
 \tilde{\rho}_L(x)+\frac{\pi}{m}\frac{1-K_0^2}{8u_0\sqrt{K_0}} 
\left[\frac{1}{\pi}\partial_x(\tilde{\rho}_L(x)\tilde{\varphi}_R(x))+\tilde{\rho}_R^2(x)\right]\,.
\end{eqnarray}
Here $\tilde{\varphi}_{\eta}(x)$ is defined by the usual relation
\begin{equation}
 \tilde{\rho}_\eta(x)=\frac{\eta}{2\pi} \partial_x\tilde{\varphi}_\eta(x)\,.
\end{equation}
In the leading order in $\rho/mv_F$ the physical density $\rho(x)=\rho_L+\rho_R \simeq \sqrt{K_0}(\tilde{\rho}_R+\tilde{\rho_L})$.

\section{Kinetic equation and chiral hydrodynamics}
\label{Supplementary2}
We  now discuss the relation  between the kinetic approach, developed above  and 
hydrodynamics description for 1D fermions with generic finite range interaction, 
developed in Ref. \cite{PGM2012}.

In terms of the bosonic densities  the  Hamiltonian of the system  can be written as
[see main text, Eq. (13)]
\begin{equation}
\label{Sup:Hboson}
H=\sum_{\eta} \int dx \left[\pi u_0\tilde{\rho}_\eta^2+\frac{2\pi^2}{3m^*} \tilde{\rho}_\eta^3\right]+
\frac12\int dx dx' \tilde{\rho}_\eta(x)V(x-x')\tilde{\rho}_\eta(x')\,,
\end{equation}
where we approximate the interaction vertex $\Gamma_{\bf q} \simeq \Gamma_{{\bf q}=0}$ 
and use real space representation.

The operators of chiral density components satisfy Heisenberg equation
\begin{equation}
\label{Supp:HydrodynamicEquation}
 \partial_t \hat{\tilde{\rho}}_\eta+\eta\left(u_0+\frac{2\pi}{m^*}\hat{\tilde{\rho}}_\eta\right)\partial_x \hat{\tilde{\rho}}_\eta
+\frac{\eta}{2\pi}\int dx' V(x-x')\partial_{x'}\hat{\tilde{\rho}}_\eta(x')
=0\,.
\end{equation}
In the  classic limit  the operators  in Eq. (\ref{Supp:HydrodynamicEquation})
are replaced by the  real density field.    
By ignoring  the difference between  density operators and their expectation values one
neglects the quantum loop corrections  to the classical equations of motion.
Such corrections play an important role in evolution of the density field Ref.\cite{PGM2012}, 
in particular in the region where hydrodynamic equations develop instabilities 
(and phase space of quasi-particle acquires an  inverse population). 
Sufficiently strong electron  interaction prevents the emerging instabilities in hydrodynamic theory, 
which  allows to neglect the loop corrections in a controlled way. 
For  the case of finite range interaction 
\begin{equation}
 g(q)=\frac{1}{l_0m}e^{-q^2l_{\rm int}^2}
\end{equation}
the hydrodynamics is  justified, provided that 
\begin{equation}
 \sqrt{\frac{l_{\rm int}^2 \Delta\rho}{l_0}}\gg1 \quad {\rm and}\quad l_0\Delta\rho\ll 1.
\end{equation}
Here $\Delta\rho$ is the amplitude of the density perturbation in the initial state. 

The  classic hydrodynamic theory can be straightforwardly derived from  
the kinetic description of the main text. 
For the right-moving particles (from now on we focus on this case  and omit the chirality index $\eta$) 
the kinetic equation reads
\begin{eqnarray}
\label{Supp:Vlasov}
\partial_t \tilde{f}(p,x,t) +\left(u_0+\frac{p}{m^*}\right)\partial _x \tilde{f}(p,x,t)  
+\int \frac{dp}{2\pi} e^{-ipy} \tilde{f}(x,y,t)
\left[\tilde{\phi}\left(x+\frac{y}{2}\right)-\tilde{\phi}\left(x-\frac{y}{2}\right)\right]=0\,,\\
\phi(x,t)=\int dx'  V(x-x')\tilde{\rho}(x',t)\,.
\end{eqnarray}

The equation (\ref{Supp:Vlasov}) should be supplied with the  initial conditions $\tilde{f}_0(x, p)$, that needs to  be calculated separately.  As in the main text, we assume that the perturbation in electronic density is created by the applying  the smooth external potential $U(x)$ to the uniform Fermi sea. 
In this case  the curvature of electronic spectrum has little effect on the initial Wigner function, and  
the standard bosonization technique 
enables us  to find  $\tilde{f}_0(x,p)$. In the vicinity of the right Fermi point (cf. discussion of the Wigner function for non-interacting fermions in Ref.\cite{PGM2012})  the Wigner function can be  written as 
\begin{equation}
\label{Supp:f0}
\widetilde{f}_0(x, p)=\int \frac{dy}{2\pi i (y-i0)}\exp\left[-ip y+2\pi i\int_{x_-}^{x_+}\tilde{\rho}_0(x')dx'\right]\,, 
\qquad x_\pm \equiv x\pm\frac{y}{2}, 
\end{equation}  
where   $\tilde{\rho}_0(x)$  is the expectation value of fermionic density  in the external potential $U(x)$. 
We note, that  the details of the interaction are  encoded in 
the static Wigner function only through $\tilde{\rho_0}$.
Several simple facts about equation (\ref{Supp:Vlasov}) help to clarify its connection to hydrodynamics.

In the limit ($m^*=\infty$) Eq. (\ref{Supp:Vlasov}) yields
\begin{equation}
\label{Supp:fLL}
\widetilde{f}(x, p, t)=\int \frac{dy}{2\pi i (y-i0)}\exp\left[-ip y+2\pi i\int_{x_-}^{x_+}\tilde{\rho}(x', t)dx'\right]\,. 
\end{equation}  
This corresponds to  density evolution 
\begin{equation}
\partial_t \tilde{\rho}+u_0\partial_x \tilde{\rho}
+\frac{1}{2\pi}\int dx' V(x-x')\partial_{x'}\tilde{\rho}(x')
=0\,
\end{equation}
in accordance   with harmonic LL model. 
As expected, Eq.(\ref{Supp:Vlasov}) is exact in the limit $m\rightarrow\infty$. 

Performing the gradient expansion in Eq.(\ref{Supp:Vlasov}) one obtains the standard Boltzmann equation
\begin{equation}
\label{Supp:VlasovReduced}
\partial_t \tilde{f}(p,x,t) +\left(u_0+\frac{p}{m^*}\right)\partial _x \tilde{f}(p,x,t)  
-\partial_x\phi(x)\partial_p\tilde{f}(x, p, t)=0\,.
\end{equation}
Approximating the initial condition  (\ref{Supp:f0}) by 
\begin{equation}
\label{Supp:f0Hyd}
 \tilde{f}_0(x, p)=\Theta(2\pi\tilde{\rho}_0(x)-p)\,.
\end{equation}

one finds the formal solution of  Eq.  (\ref{Supp:VlasovReduced}) 
\begin{equation}
\tilde{f}(x, p, t)=\Theta(2\pi\rho(x, t)-p)\,,
\end{equation}
 where the density $\rho(x, t)$ satisfies  the hydrodynamic equation (\ref{Supp:HydrodynamicEquation}).

\section{Numerical solution of the kinetic equation}
\label{Supplementary3}
In this section we briefly discuss  the  algorithm used for numeric simulation of Eq. (\ref{Supp:Vlasov}). 
We use   the model of  fermions  on a ring,   of  the  circumference $L_x$. 
This induces periodic  boundary conditions for the Wigner function $f(x, y)$ 
with the period with period $L_y=2L_x$, as a function of 
$y$ and $x$ correspondingly.
The fermionic momentum $p$ in $f(x, p)$ is quantized in units of $2\pi/L_y$, while the momentum $q$ conjugate to $x$ is quantized in units of $2\pi/L_x$. To perform numerical  simulations we impose the cut-off  $2\pi N_x/L_x$ and $2\pi N_y/L_y$ for momenta $q$ and $p$ 
respectively.
In our calculations,  the values of the parameters $L_x=4000$ (in units where $\lambda_F\equiv mV_F=2\pi$),   $N_x\sim 2500$ and $N_y\sim500$ were used. 
We checked that the final results are  stable with respect to the variation of these parameters.
We model the  initial density bump  by a Gaussian with the dispersion $\sigma=200$ that contain  
$N\approx 5$ particles.

Periodic boundary  conditions enable the use of  fast Fourier transform algorithm for the  calculation
of $\partial_t \tilde{f}$, given by   (\ref{Supp:Vlasov}). 
Combined with the standard fourth-order Runge-Kutta time stepper this provides us with the fast and accurate algorithm for the numerical solution of Eq. (\ref{Supp:Vlasov}).


\begin{thebibliography}{99}
\bibitem{3-body-interaction}
The three body interaction turns out to be important for the calculation of the inelastic relaxation time and will be discussed 
in details in the separate publication.
\end{thebibliography}
\end{document}